\documentclass[mathleft,fleqn,%
]{an}
\usepackage{graphicx}
\usepackage[varg]{txfonts}
\usepackage{natbib}
\bibpunct{(}{)}{;}{a}{}{,}
\begin{document}

\Pagespan{1}{}
\Yearpublication{2014}%
\Yearsubmission{2014}%
\Month{0}%
\Volume{999}%
\Issue{0}%
\DOI{asna.201400000}%

\title{Testing Asteroseismic Scaling Relations \\using Eclipsing Binaries in Star Clusters and the Field}

\author{K. Brogaard\inst{1}\fnmsep\thanks{Corresponding author:
        {kfb@phys.au.dk}}
\and J. Jessen-Hansen\inst{1}
\and R. Handberg\inst{1}
\and T. Arentoft\inst{1}
\and S. Frandsen\inst{1} 
\and F. Grundahl\inst{1}
\and H. Bruntt\inst{1}
\and E.\,L. Sandquist\inst{2}
\and A. Miglio\inst{3}
\and P.\,G. Beck\inst{4}
\and A.\,O. Thygesen\inst{5}
\and K.\,L. Kj\ae rgaard\inst{1}
\and N.\,A. Haugaard\inst{1}
}
\titlerunning{Testing Asteroseismic Scaling Relations}
\authorrunning{K. Brogaard et al.}
\institute{
Stellar Astrophysics Centre, Department of Physics and Astronomy, Aarhus University, Ny Munkegade 120, DK-8000 Aarhus C, Denmark
\and 
Department of Astronomy, San Diego State University, USA
\and 
School of Physics and Astronomy, University of Birmingham, United Kingdom
\and 
Service d'Astrophysique, CEA, Paris-Saclay, France
\and
Zentrum f{\"u}r Astronomie, Landessternwarte, Universit{\"a}t Heidelberg
}

\received{XXXX}
\accepted{XXXX}
\publonline{XXXX}


\abstract{%
The accuracy of stellar masses and radii determined from asteroseismology is not known! \\
We examine this issue for giant stars by comparing classical measurements of detached eclipsing binary systems (dEBs) with asteroseismic measurements from the {\it Kepler} mission.
For star clusters, we extrapolate measurements of dEBs in the turn-off region to the red giant branch and the red clump where we investigate the giants as an ensemble.
For the field stars, we measure dEBs with an oscillating giant component.
These measurements allow a comparison of masses and radii calculated from a classical eclipsing binary analysis to those calculated from asteroseismic scaling relations and/or other asteroseismic methods. 
Our first results indicate small but significant systematic differences between the classical and asteroseismic measurements. In this contribution we show our latest results and summarize the current status and future plans.\\
We also stress the importance of realizing that for giant stars mass cannot always be translated to age, since an unknown fraction of these evolved through a blue straggler phase with mass transfer in a binary system. Rough estimates of how many such stars to expect are given based on our findings in the open clusters NGC6819 and NGC6791.
}

\maketitle

\section{Introduction}

The space missions {\it Kepler} and CoRoT have opened up for large scale applications of asteroseismology - the study of stars through their oscillations. We are entering into a new era of precision asteroseismology where parameters of stars, like masses, radii and ages, can be determined for single stars to a significantly higher precision than before. However, precision is not necessarily accuracy, and therefore it is of importance to test the accuracy level of asteroseismic measurements with independent methods. 

While the radii can be tested for nearby single stars with accurate distances or interferometric radii \citep{Silva2013} the only way to compare asteroseismic mass measurements to an accurate model-independent mass is using detached eclipsing binaries (dEBs). We are in the process of building up a set of measurements of dEBs for testing the accuracy of asteroseismic measurements \citep{Brogaard2015}. One focus area is star clusters where we can extrapolate measurements of dEBs in the turn-off region to the red giant branch and the red clump where we investigate the giants with ensemble asteroseismology. We also study dEBs in the field that have an oscillating red giant component, so that the asteroseismic measurements can be directly compared to the more classical measurements from the eclipsing binary analysis.

So far we are concentrating on testing the asteroseimic scaling relations for mass and radius of giant stars. These relations, introduced elsewhere in these proceedings, use the global asteroseismic parameters known as the large frequency separation $\Delta \nu$ and the frequency of maximum power $\nu_{\rm max}$ and a non-seismic measurement of the effective temperature $T_{\rm eff}$. For many stars these are the only data which will be available because the stars are too faint to allow extraction of more information even when observed with {\it Kepler} or CoRoT. Therefore, we have to be able to trust the asteroseismic scaling relations in order to correctly interpret inferences on distant stars in the Galaxy and to avoid situations where the asteroseismic analysis is questioned when the outcome is not what was expected \citep{Epstein2014}. 

\section{Star clusters}

There are four open clusters in the {\it Kepler} field of view. The two oldest and most crowded ones (NGC6791 and NGC6819) are observed by {\it Kepler} in terms of so-called super-stamps that are large regions centered on the clusters. For the two younger and less populated clusters, only individually selected targets are observed. From these data we have identified a large number of detached eclipsing binary members of NGC6791 \citep{Brogaard2016b}, and a smaller number in NGC6819 \citep{Jeffries2013,Sandquist2013} and NGC6811 \citep{Sandquist2016}. The few candidates we identified in NGC6866 were found to be non-members. 

In previous works we compared the masses of red giants in NGC6791 and NGC6819 derived by isochrone extrapolation from eclipsing binaries at the turn-off to those derived from asteroseismic scaling relations using {\it Kepler} observations of giant stars \citep{Brogaard2012,Sandquist2013,Brogaard2015}. In those cases we found that a small, but significant downward correction to the $\Delta \nu$ scaling relation is needed in order to obtain agreement with dEB measurements. In these clusters the giant stars on the red giant branch (RGB) and those in the red clump (RC) were also found to need a different relative correction in order have identical distances \citep{Miglio2012,Brogaard2015,Handberg2016}. These findings are all in agreement with theoretical predictions for corrections to the scaling relations \citep{Miglio2013}.

In NGC6791, the dEB V9 \citep{Kaluzny1993} was found to be a member with an early giant component and companion in the turn-off phase. Measurements of the masses of this system will decrease the range over which the isochrone needs to be extrapolated to compare to the only slightly brighter red giants that have measured oscillations. The measurement of masses of this system is complicated by the fact that the giant is magnetically active due to a short orbital period of 3.187 days. This causes significant out-of-eclipse variations and periodic features in the spectral lines. Despite this, it seems that we will be able to reach a precision and accuracy on the mass measurements below 2\% through a careful analysis \citep{Brogaard2016b}. Our preliminary measurement yields $M_{\rm V9p}=1.14\pm0.02 M_{\odot}$ in perfect agreement with our previous estimate extrapolated from the turn-off dEBs V18 and V20 \citep{Brogaard2012} and enforcing the suggestion that the uncorrected asteroseismic scaling relations overestimate masses slightly.

The young open cluster NCG6633 has been observed by the CoRoT mission and three oscillating giant members have been identified \citep{Barban2013}. The investigation by \cite{Lagarde2015} compared radii of the these stars calculated from uncorrected asteroseismic scaling relations to those calculated using distances from the Hipparcos catalog \citep{Leeuwen2007, Leeuwen2009}. They found a potential disagreement but concluded that it was likely not significant. However, the corrections to the asteroseismic scaling relations calculated by Miglio et al. (2013;Fig.2) suggest very significant upwards corrections to the $\Delta \nu$ scaling for such massive core-helium burning stars. This would enhance the discrepancy found by \cite{Lagarde2015} thus questioning the validity of the corrections to the scaling relations for helium-burning giants with masses around $3 M_{\odot}$. But this of course relies on the accuracy of the Hipparcos distance to NGC6633. We are in the process of obtaining observations of a dEB at the turn-off in this cluster in order to investigate the apparent discrepancy.

\subsection{Over-massive stars}

As a by-product of the ensemble asteroseismology in the open clusters NGC6791 and NGC6819 we found giant stars that are over-massive compared to the ensemble \citep{Brogaard2012,Brogaard2015,Corsaro2012,Handberg2016}. We expect those stars to have evolved through a blue straggler phase with mass transfer in a binary system. This means that the present mass cannot be translated into an age as can be done for giant stars that evolved a single stars. While this is not a problem when the star is a cluster member where the age can be determined from the other stars, it would cause severe complications for age estimation if the star was found in the field. As an example, we mention that the young and alpha-rich stars found by \cite{Chiappini2015,Martig2015} could in principle be old stars that experienced mass-transfer during their evolution. 

Our asteroseismic investigation of NGC6819 \citep{Handberg2016} found a fraction of over-massive giant members of more than 10\% in this cluster (5 clear detections and another 6 to be investigated further at the time of writing). For NGC6791 the proper motion study of \cite{Platais2011} found a number of bright giants outside the predicted single star evolution sequence that they suggested belong the the horizontal branch. However, it was found by \cite{Brogaard2012} that the bluer stars in this sample were too faint to belong to the horizontal branch while the redder part would likely belong to a group of stars that evolved through a blue straggler phase due to their positions in the colour-magnitude diagram that were otherwise not easily explainable. These investigators also identified an additional star as belonging to the group of over-massive stars \citep{Brogaard2012}. The radial-velocity study of bright stars in NGC6791 by \cite{Tofflemire2014} revealed that many of the stars suggested by \cite{Platais2011} to be members in non-standard evolutionary phases were in fact non-members. Only three of the redder stars retained their cluster membership. We have conducted a preliminary investigation of these stars and found that two of them are over-massive compared to the ensemble while the third might be blended in the photometry of \cite{Platais2011}, as this star sits right in the RC in the photometry of \cite{Stetson2003}. All in all we find three giant stars that are over-massive compared to the ensemble of 60 stars \citep{Corsaro2012}. While this is not a complete sample, indications are that about 5\% of the giants are over-massive in this cluster, taking into account that two of the over-massive stars were not it the sample of \cite{Corsaro2012}. This is not necessarily an indication that fewer stars evolve through non-standard evolution in this cluster compared to NGC6819 since NGC6791 also hosts a group of extreme horizontal branch stars. 

For more massive stars in younger star clusters our statistics is not as good. In NGC6811 only 6 giant members have been identified with asteroseismology \citep{molenda2014, Corsaro2012, Hekker2011, Stello2011} and non of them are found to be over-massive compared to the ensemble. This only puts a very week limit of $\lesssim14\%$ over-massive stars in this mass range. Adding asteroseismic ensemble studies from more open clusters in the K2 fields will give us better statistics in the future. 

With the current information indications are that about 5-10\% of stars in open clusters are over-massive stars that underwent non-standard evolution that prevents estimates of age from their masses. Whether or not a similar number is to be expected for field stars depends on whether binary fractions are significantly different in the field compared to open clusters. An additional complication is that most of the over-massive stars identified in NGC6819 were found to belong to very long period binary systems while those in NGC6791 seem to be single stars. Therefore it is not clear how the percentages will change for samples of stars where binarity can be ruled out. But the potential appearance of over-massive stars arising from non-standard evolution is certainly something that should be considered when interpreting asteroseismic ages for large populations of giant stars.  

\section{Field stars}
We are in the process of observing and analysing several detached eclipsing binaries with a red giant component showing solar-like oscillations. These were identified in the {\it Kepler} eclipsing binary catalog \citep{Prsa2011}. From those that were found to show solar like oscillations and a total eclipse several days long in the {\it Kepler} light curve we selected the ones that were most likely to be SB2 binaries for spectroscopic follow-up with the FIES spectrograph at the Nordic Optical Telescope and/or the HERMES spectrograph at the Mercator telescope, both located at the Observatorio del Roque de los Muchachos on La Palma.

\begin{figure*}
\includegraphics[width=17cm]{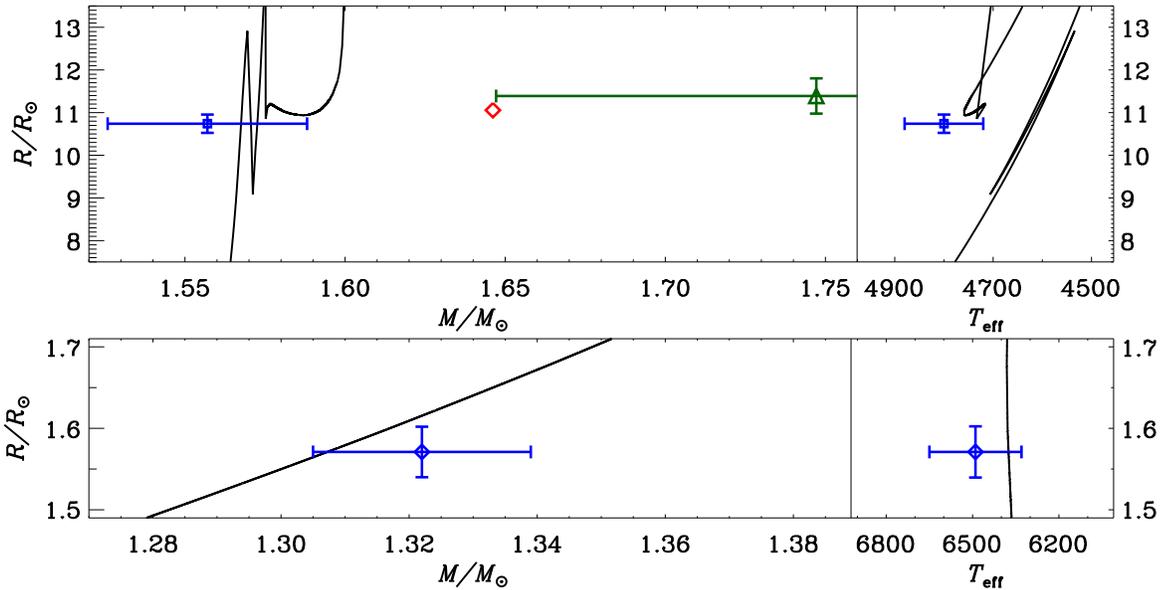}
\caption{Mass--radius and $T_{\rm eff}$--radius diagrams for the eclipsing binary KIC8410637. Blue points are the masses and radii as determined from the eclipsing binary analysis. Then green triangle is the mass and radius of the giant component as determined using the asteroseismic scaling relations. Error-bars are $1-\sigma$. The red diamond is the asteroseismic measurement when using theoretical corrections to the scaling relations from \cite{Miglio2013} assuming the giant is in the RGB phase of evolution. The black line is a PARSEC isochrone for our measured metallicity and an age of 2.6 Gyr.} 
\label{mr8410637}
\end{figure*}

\begin{figure*}
\includegraphics[width=17cm]{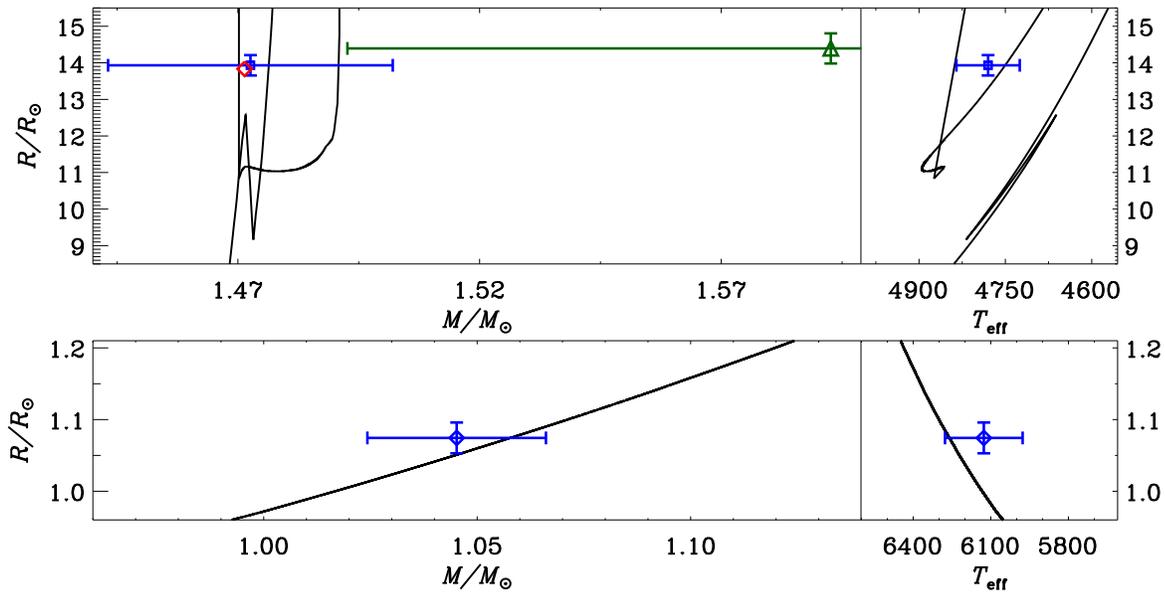}
\caption{As Fig.~\ref{mr8410637} but for the eclipsing binary KIC9540226. The PARSEC isochrone is for our measured metallicity and an age of 2.6 Gyr.}
\label{mr9540226}
\end{figure*}

We finished the eclipsing binary analysis of the first of these systems, KIC8410637, which is published in \cite{Frandsen2013}. The second system, KIC9540226, is still under our analysis and the details will be published elsewhere \citep{Brogaard2016}. Two studies provided partial analysis of this system; \cite{Gaulme2014} did a light curve analysis of the binary orbit and measured global seismic parameters while \cite{Beck2014} published their estimates of the global seismic parameters along with a spectroscopic orbit determined from radial velocities of the primary component only. From the same data we have now measured also the radial velocities of the secondary which allowed us to derive precise masses and radii. Additional HERMES spectra were obtained during the total eclipse for the purpose of $T_{\rm eff}$ and abundance analysis \citep{Frandsen2013b} but the S/N was not high enough for optimal results so we ended up basing that part of the analysis on disentangled spectra using all HERMES spectra. We include our preliminary results here to add to the discussion of the asteroseismic scaling relations. The parameters derived for the giant in each of the two systems can be found in table~\ref{tab1} and mass-radius diagrams are shown in Figs.~\ref{mr8410637} and ~\ref{mr9540226}. The fact that the two systems are found to have very similar ages is either a coincidence or due to a selection effect that we have not worked out yet. 

\begin{table}
\centering
\caption{Parameters of the giant components of the dEBs}
\label{tab1}
\begin{tabular}{lcc}\hline
Property & KIC8410637 & KIC9540226\\ 
\hline
Mass $\rm (M_{\odot})$ & $1.557\pm0.028$ & $1.473\pm0.029$\\
Radius $\rm (R_{\odot})$ & $10.74\pm0.11$ & $13.93\pm0.28$\\
$T_{\rm eff}$ (K) & $4800\pm80$ & $4780\pm55$\\
log$g$ & $2.57\pm0.01$ & $2.32\pm0.01$\\
$[\rm Fe/H]$ & $+0.08\pm0.13$ & $-0.21\pm0.10$\\
age (Gyr)	& $2.6\pm0.2$ & $2.6\pm0.2$\\
\hline
\end{tabular}
\end{table}

Fig.~\ref{mr8410637} compares the measurements of KIC8410637 from our eclipsing binary analysis to measurements of the giant component using the asteroseismic scaling relations and to a PARSEC isochrone \citep{Bressan2012}. On the mass--radius plane the isochrone matches both components from the eclipsing binary analysis well within their 1$-\sigma$ error-bar, but the spectroscopically measured $T_{\rm eff}$ of the giant star is only consistent with the isochrone if the star is in the red clump (RC) phase of evolution, burning helium in its core. While asteroseismic methods have been developed for distinguishing between the red giant branch (RGB) and RC phases of evolution \citep{Bedding2011,Kallinger2012} the oscillation spectrum of KIC8410637 is difficult to interpret and no clear-cut conclusions about this has yet been published. According to S. Frandsen (private comm.), the measured g-mode period spacing suggests that the star is in the RC phase consistent with what is suggested by the relatively high $T_{\rm eff}$. On the other hand, if the giant is now in the RC phase it previously went through the helium flash at the tip of the red giant branch. According to PARSEC models the star would have been so large around that point that the secondary would have orbited inside its atmosphere for part of its very eccentric orbit \citep{Frandsen2013}, and it is difficult to imagine how the secondary could have survived that.

As can be seen in Fig.~\ref{mr8410637}, the mass and radius of the red giant differ significantly when the asteroseismic scaling relations are used without corrections, while they are in agreement well within the asteroseismic error-bar when using a correction based on \cite{Miglio2013}. However, this correction assumes that the giant star in on the RGB burning hydrogen in a shell. If the star is instead in the red clump (RC) phase of evolution, burning helium in its core, the predicted correction is of similar size but in the opposite direction, which would cause a discrepancy close to the 3$-\sigma$ level. This could either be taken as evidence that the giant is in the RGB phase or that we are not able to predict corrections to the asteroseismic scaling relations properly. The evolutionary state of the star needs to be firmly established before we will know. Hopefully this can be established through a detailed asteroseismic modeling of individual frequencies.

The mass--radius diagram of KIC9540226 in Fig.~\ref{mr9540226} shows that in this case there is agreement between the mass and radius of the giant as found from the eclipsing binary analysis and the asteroseismic scaling relations even without a correction. Applying the correction suggested by \cite{Miglio2013} under assumption that the giant is on the RGB makes the agreement even better. However, when comparing to the isochrone it is found that also in this case the spectroscopic $T_{\rm eff}$ is only consistent if the star is in a helium burning phase -- this time on the early asymptotic giant branch burning Helium in a shell. If so, that would again mean that the correction to the scalings should be in the opposite direction, creating significant disagreement with the classical measurements. However, it would have been even more difficult for this system to have survived the evolution through the upper RGB, since the orbital period, and therefore also the distance between the stars is much smaller in this case than for KIC8410637, and therefore the components would have been in deep contact. This suggests that the giant in KIC9540226 is on the RGB and that our measured $T_{\rm eff}$ is too high or  the $T_{\rm eff}$ predicted by the PARSEC models is too low. If we assume that this systematic is the same for the case of KIC8410637 then that system is also likely to be in the RGB phase and the corrections to the asteroseismic scaling relations seem to be working well. We hope to obtain photometric $T_{\rm eff}$ estimates of the systems and measure their reddening from the interstellar Na lines \citep{Munari1997} to investigate this further.

Despite the complications it seems that the asteroseismic scaling relations including a correction according to \cite{Miglio2013} are doing quite well. We note especially the extreme agreement for log$g$ between the classical and asteroseismic measurements, at the level of 0.01 dex, which is the uncertainty of the asteroseismic measurement. This suggests that at least for the observed masses and metallicities the scaling of $\nu_{\it max}$, which is the only asteroseismic parameter that enters the equation for log$g$, is quite accurate. Therefore, log$g$ values derived from the asteroseismic scaling relations can be trusted to within their uncertainty and the asteroseismic log$g$ values should be exploited in spectroscopic analysis of stars where uncertainties in log$g$ are often about an order of magnitudes larger and log$g$ correlates with other parameters.

\section{Conclusions and outlook}

Our on-going investigations of the asteroseismic scaling relations suggest that they are accurate to within their uncertainties for giant stars as long as corrections to $\Delta \nu$ are calculated and applied along the lines of \cite{Miglio2013}. The fact that asteroseismic log$g$ values are extremely consistent with our independent measurements implies that the scaling for $\nu_{\rm max}$ is reliable.
Consistency among the different formulations of the scaling relations for stars in clusters with independent distance estimates, once an expected correction to $\Delta \nu$ has been applied, also suggests that no correction is needed for $\nu_{\rm max}$.
 
In some cases there are suggestions that discrepancies could be present, but at the moment these cannot be claimed to be due to the scaling relations, since they could also be caused by systematic uncertainties in our spectroscopic $T_{\rm eff}$ scale, the temperature scale of the PARSEC stellar models, and the Hipparcos distance to NGC6633. 

From our investigations we find that a fraction of 5-10\% of stars in open clusters are over-massive and emphasize the need to consider the potential presence of such stars when interpreting asteroseismic analysis of large populations of stars. Unfortunately we do not know whether the fraction of over-massive giants is the same for field stars and open clusters. 

The surface gravity determined from asteroseismic scaling relations is found to be identical to that derived from the eclipsing binary analysis to well within the asteroseismic uncertainty estimate, even when that uncertainty reaches as low as 0.01 dex. Since this is about an order of magnitude better than is usually derived from a spectroscopic analysis we advocate the use of asteroseimically determined log$g$ values in spectroscopic analysis to minimize errors caused by correlations between log$g$ and $T_{\rm eff}$.

We will continue to investigate these issues in more detail and extend our analysis to more dEBs and star clusters observed by the K2 mission. This will expand the mass and metallicity range where we can have faith in the accuracy of the asteroseismic scaling relations.


\acknowledgements
 Funding for the Stellar Astrophysics Centre is provided by The Danish National Research Foundation (Grant agreement no.: DNRF106). The research is supported by the ASTERISK project (ASTERoseismic Investigations with SONG and Kepler) funded by the European Research Council (Grant agreement no.: 267864). 
 KB acknowledges funding from the Villum Foundation.
  

%
%

\bibliographystyle{an} 


\end{document}